\documentclass{moriond}

\def\ie{{\it i.e.}}
\def\eg{{\it e.g.}}
\def\etc{{\it etc}}

\def\to{\rightarrow}

\def\be{\begin{equation}}
\def\ee{\end{equation}}
\def\bea{\begin{eqnarray}}
\def\eea{\end{eqnarray}}

\newcommand{\Photo}{}

\begin{document}
\vspace*{4cm}

\title{Dark Photons, Kinetic Mixing and Light Dark Matter From 5-D
}
\author{ Thomas G. Rizzo }

\address{SLAC National Accelerator Laboratory,\\
2575 Sand Hill Rd., Menlo Park, CA 94025 USA}

\maketitle\abstracts{
Extra dimensions provide a unique tool for building new physics models. Here we extend the kinetic mixing/dark photon mediator scenario for the case of complex scalar dark matter interacting 
with the Standard Model to 5-D. We assume that the inverse size of the new, flat extra dimension is $\sim 10 - 1000$ MeV, the mass range of interest in numerous current experiments, and discuss 
the resulting phenomenology. Here we see that 5-D constructions can be used to soften some of the possible tuning issues which are sometimes encountered in the corresponding 4-D models.}

\section{Introduction and Overview}

Although we believe dark matter (DM) exists we actually know very little about its true nature. With the lack of any clean WIMP signatures coming from DM detectors or the LHC, it behooves us to 
think more broadly, both theoretically and experimentally, and this has thrown the doors wide open to many new and interesting possibilities. One scenario which has gotten much attention 
recently is that of DM interacting with the Standard Model (SM) through the kinetic mixing (KM) portal, specifically, with the DM and the corresponding dark photon (DP) gauge field both typically 
having similar masses in the $\sim$ 10 -1000 MeV range. In the simplest realization, the only other parameter besides these two masses and the dark gauge coupling is $\epsilon$,  
which describes the strength of this KM. Typical values of $\epsilon \sim \rm O(10^{-(3-4)})$ can lead to the the observed DM relic density via thermal freeze-out while still satisfying other 
experimental constraints. In looking beyond the usual frameworks, extra dimensions (ED) can be very useful as a model building tool. Here we will consider a very simple extension of the KM portal 
model to the case of 5-D with the inverse size of the additional, flat ED in the mass range above implying the existence of Kaluza-Klein excitations at this scale\cite{big}. This being the case, the SM 
must not directly experience this ED and so is confined to a brane at one end of this ED interval with only the DP and the SM singlet DM in the bulk. For DM in this mass range, CMB and 21 cm 
constraints tell us that the DM must have a p-wave annihilation cross section which is most easily accomplished when the DM mass is below that of the (lightest) mediator and when the DM 
is a complex scalar, although there are several other possibilities. Note that the DM must be a {\it complex} scalar so that it can have a dark charge allowing it to couple directly to the $U(1)_D$ 
DP gauge field. In our discussion we will ignore the possibility where the scalar DM field gets a vev for simplicity although this is certainly possible. We find that the introduction of an ED can help 
to explain some of the aspects of the 4-D scenario that are usually put in by hand or that require some fine-tuning. As will be seen below, these ED scenarios require the existence of brane-localized 
kinetic terms (BLKTs) on one or the other branes in order to obtain the correct phenomenology and additional model-building flexibility.

\section{Analysis Survey}

To be concrete, the basic setup we consider has an ED living on a brane-bounded interval $0\leq y\leq \pi R$ and is described by the following action:
\begin{equation}
S=S_1+S_2+S_{BLKT}\,
\end{equation}
where the various pieces of the action are given by 
\begin{equation}
S_1=\int ~d^4x ~\int_{0}^{\pi R} ~dy ~\Big[-\frac{1}{4} \hat V_{AB} \hat V^{AB}  ~\Big(-\frac{1}{4} \hat B_{\mu\nu} \hat B^{\mu\nu} 
+\frac{\epsilon_5}{2c_w} \hat V_{\mu\nu} \hat B^{\mu\nu}  + L_{SM} \Big) ~\delta(y-\pi R) \Big] \,,  
\end{equation}
describing the brane-localized SM plus pure 5-D bulk gauge fields and includes the KM of $U(1)_D$ with hypercharge on the 4-D brane. The hatted fields must undergo field redefinitions to bring 
$S_1$ into canonical form and, as usual,  we define $D_A=\partial_A +ig_{5D} Q_D \hat V_A$ as the $U(1)_D$ dark gauge covariant derivative in obvious notation.  When the $\hat B$ is shifted 
to remove the KM, $\hat B \to B+\frac{\epsilon_5}{c_w} V$,  a {\it negative} BLKT is generated for the DP field which leads to tachyons/ghosts in its Kaluza-Klein (KK) spectrum. Thus we must 
add a {\it positive} gauge BLKT to (at the very least) offset this from the beginning and for later model building purposes we also will introduce a similar BLKT for the DM scalar:
\begin{equation}
S_{BLKT}=\int ~d^4x \int_{0}^{\pi R} dy ~\Big[-\frac{1}{4}  V_{\mu\nu}  V^{\mu\nu}  \cdot\delta_AR~\delta(y-\pi R)  +~(D_\mu S)^\dagger (D^\mu S)\cdot \delta_SR~\delta(y) \Big]\,
\end{equation}
with $\delta_{A,S}$ being dimensionless O(1) BLKT parameters; note the BLKTs are on opposite branes for the reasons we'll see below. The kinetic and potential pieces for $S$ are given by
\begin{equation}
S_{2}=\int ~d^4x ~\int_{0}^{\pi R} ~dy ~\Big[(D_A S)^\dagger (D^A S) +\mu_S^2 S^\dagger S -\lambda_S (S^\dagger S)^2+~\lambda_{HS} H^\dagger H S^\dagger S~\delta(y-\pi r)\Big]  \,
\end{equation}
with $H$ the SM Higgs field. Since $S$ gets no vev, we will ignore the pure $S$ potential terms for simplicity in the discussion below, $\eg$, for convenience we will set the bulk mass 
of $S$ to zero in what follows although this isn't a necessary choice. 
The last term generates potentially dangerous $H$ decays after SM SSB and we will return to it shortly. Note that we have {\it not} added a dark Higgs in the bulk for SSB to generate the 
DP mass; as we'll see it is not needed. The 5-D fields $V,S$ can now be KK expanded as (suppressing indices) $V(x,y)=\sum_n v_n(y) V_n(x)$, and similarly for $S$ with $v_n\to s_n$, \etc, and 
we can define the set of quantities $\epsilon_n=\epsilon_5v_n(\pi R)$ as the KM parameters for the various KK DP tower fields. These KK modes will be simple superpositions of sines and 
cosines with the unknowns determined by the boundary conditions and overall normalizations as usual; below for simplicity we will work in the unitary/physical $V^5=0$ gauge. 
After the `undoing' of the KM by the field redefinitions, the $V$ KK tower members will mass-mix with the $Z$ through their KM-induced couplings to the SM Higgs. Once this mass matrix is 
diagonalized it results in slight shifts the $Z$ mass and couplings away from their SM values. However, when the KM parameters $\epsilon_n$  are sufficiently small (as here) this poses 
no threat to the agreement of the SM predictions with the EWK precision measurements. 

\begin{figure}
\centerline{\includegraphics[width=3.5in,angle=90]{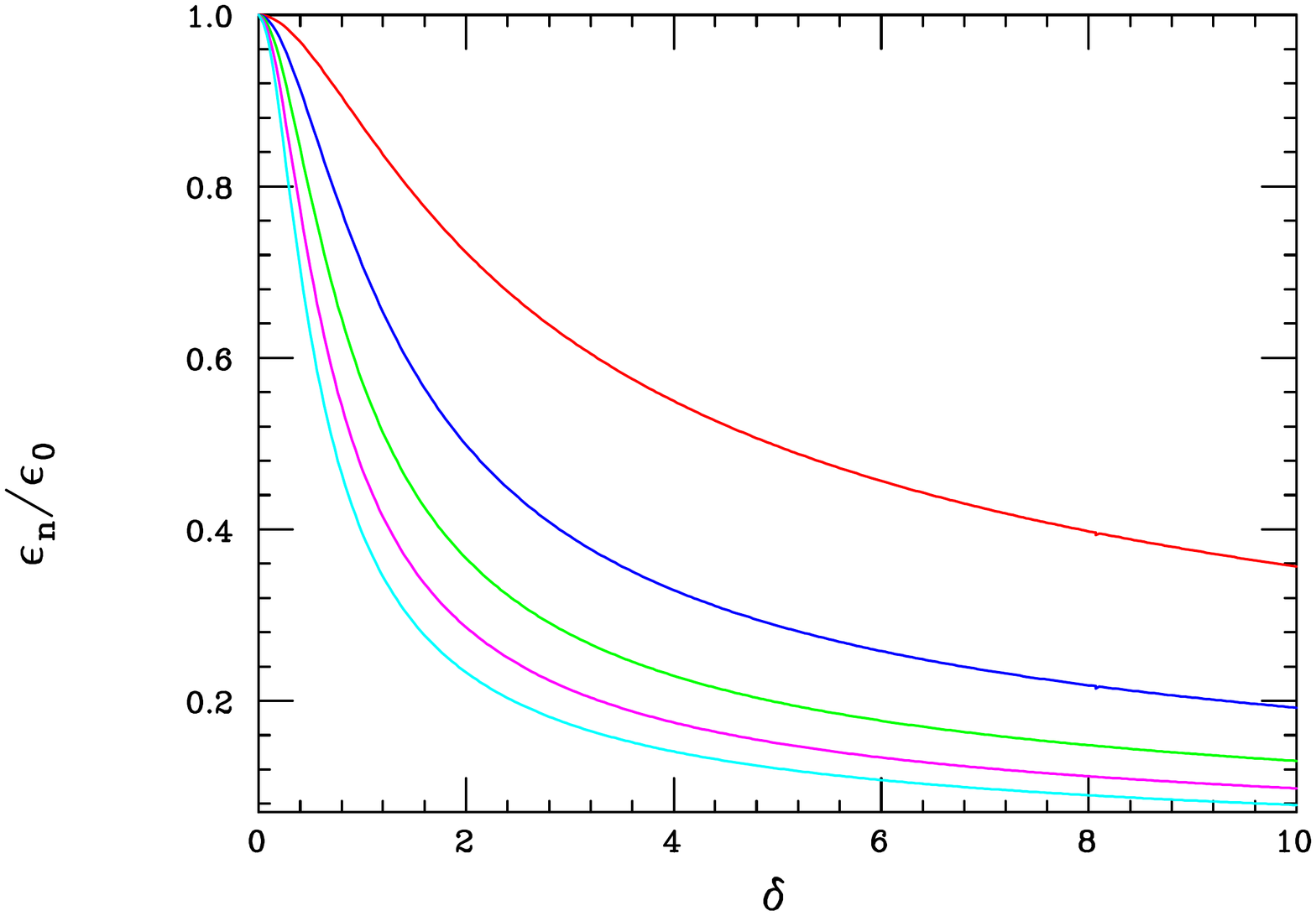}
\hspace {-2.4cm}
\includegraphics[width=3.5in,angle=90]{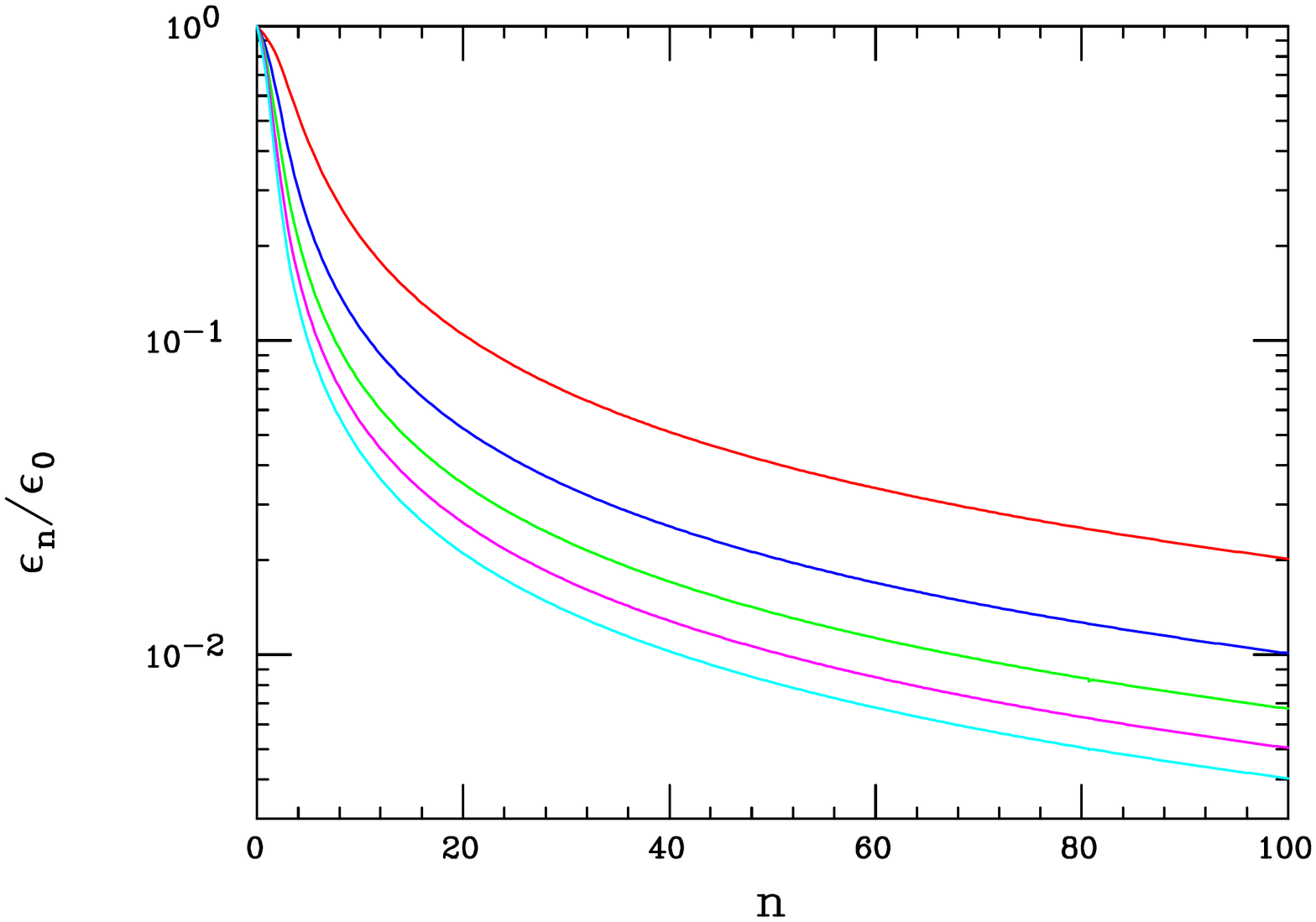}}
\vspace*{-1.5cm}
\centerline{\includegraphics[width=3.5in,angle=90]{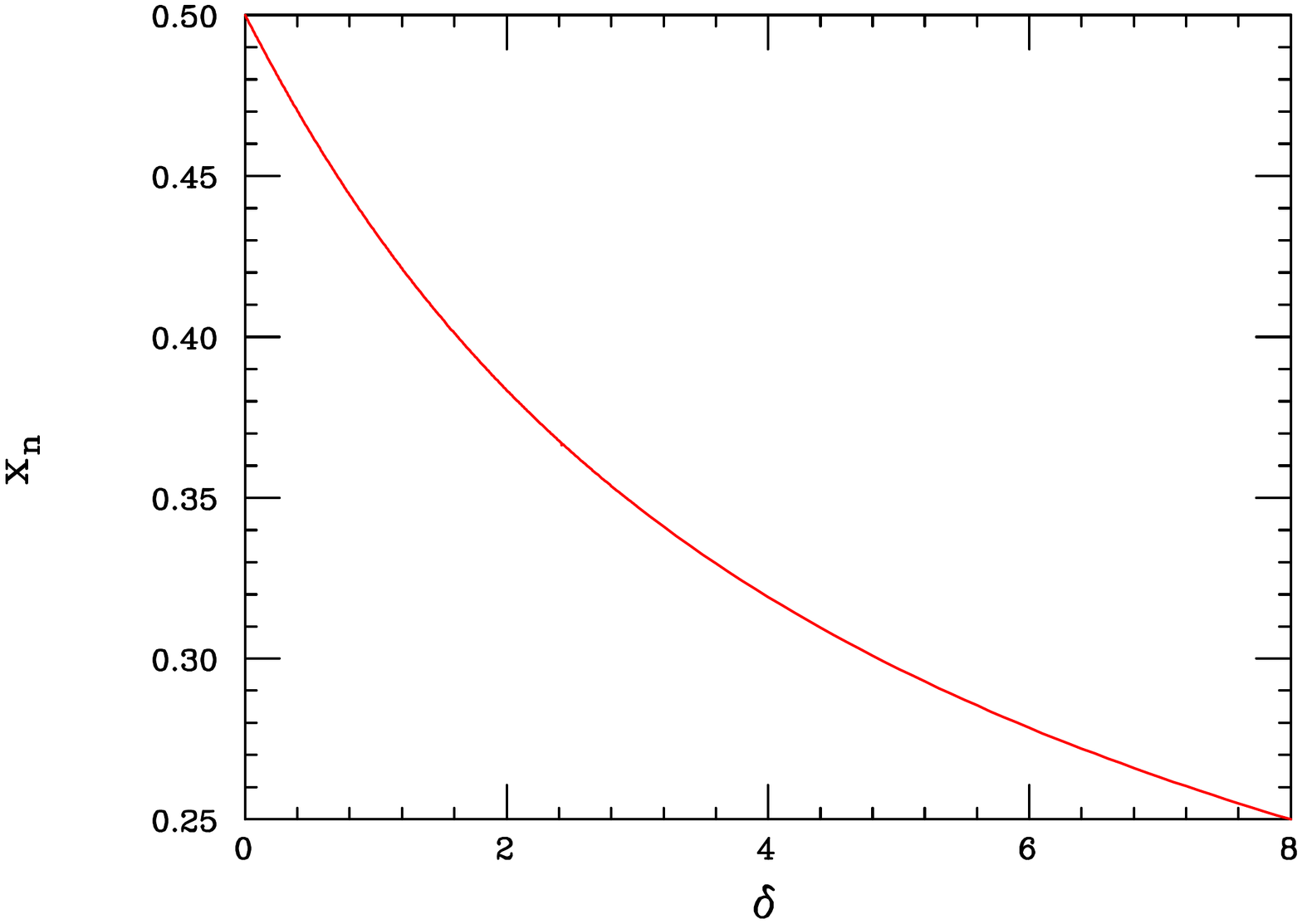}}
\vspace*{-1.30cm}
\caption{(Top left) $\delta_A$ dependence of $\epsilon_n$ for $n=1,2,..5$ from top to bottom and (Top right) $n$ dependence of $\epsilon_n$ for $\delta_A=1,2,..5$ from top to bottom. 
(Lower) The lowest root as a function of $\delta_{A,S}$. }
\label{fig1}
\end{figure}

Moving along, we choose the boundary conditions (BCs) $v_n(0)=s_n(\pi R)=0$ and the corresponding discontinuity equations for the partial derivatives due the BLKTs at the brane locations 
when solving the equations of motion. Note that if the DM had been confined to the brane opposite the SM we could not chose these BCs and, as we will see, BCs could not be used to break 
the gauge symmetry. With these specific BCs, only the BLKTs introduced above can be physically relevant as any introduced on the opposite branes will not influence the equations of 
motion for the KK states since the fields vanish there. We then 
find the masses of the KK states to be given by  $m_{V_n}R=x_n^V$ where $\cot \pi x^V_n= \frac{\delta_A}{2} x^V_n$ and similarly for $V\to S$.  A short analysis shows that this setup 
accomplishes a number of interesting things: ($i$) $x^V_1\neq 0$, \ie, there are no massless gauge modes as the gauge symmetry has been broken without the introduction of a dark Higgs 
(unlike in 4-D) by the BCs with the $V^5$ playing the role of the Goldstones. ($ii$) The masses of lightest DP and DM KK excitations are naturally of the same size without any tuning (unlike 
in 4-D)  and with the phenomenological requirement $m_{DM=S_1}>m_{DP=V_1}$ resulting from simply choosing $\delta_S>\delta_A$. In fact, requiring any specific value of the mass ratio 
$\lambda=m_{DM}/m_{DP}$, for a given value of $\delta_A$ the corresponding required value of $\delta_S$ is just given by $\delta_S = \frac{2\cot \pi\lambda x^V_1}{\lambda x^V_1}$.  
($iii$) The potentially dangerous $HS$ term in the action above vanishes by our BC choice for all KK modes of $S$; this term can't be removed by any symmetry in 4-D and so the coefficient 
of this term in the potential must be fine-tuned there.  We also find that ($iv$) the KM parameters $\epsilon_n$, as determined from the normalization of the $v_n$ in the presence of the BLKT, 
are functions only of the  $\delta_A$ and decrease rapidly in magnitude as either $n$ or $\delta_A$ are increased, \ie, $\epsilon_n^{-1} \sim 1+\Big(\frac{\delta_A x^V_n}{2}\Big)^2+
\frac{\delta_A}{2\pi}$. This implies that the heavy KK modes generally decouple from physical processes. This is shown in Fig.~\ref{fig1}; this Figure also explicitly shows how the lowest 
root for either KK tower decreases as a function of value of its associated BLKT. 
Whether or not the value of $\lambda<1/2$ strongly influences the model phenomenology. When $\lambda<1/2$ it is easy to convince oneself that if any of the KK states in either tower 
are produced they will eventually cascade decay down to stable DM as $V_1 \to SS^\dagger$ so that only a missing energy (ME) signal is the result. If $\lambda>1/2$, we have instead 
$V_1\to e^+e^-$ so that cascades can produce complex decays which will include both missing energy as well as $e^+e^-$ pairs.

\begin{figure}[htbp]
\centerline{\includegraphics[width=3.5in,angle=90]{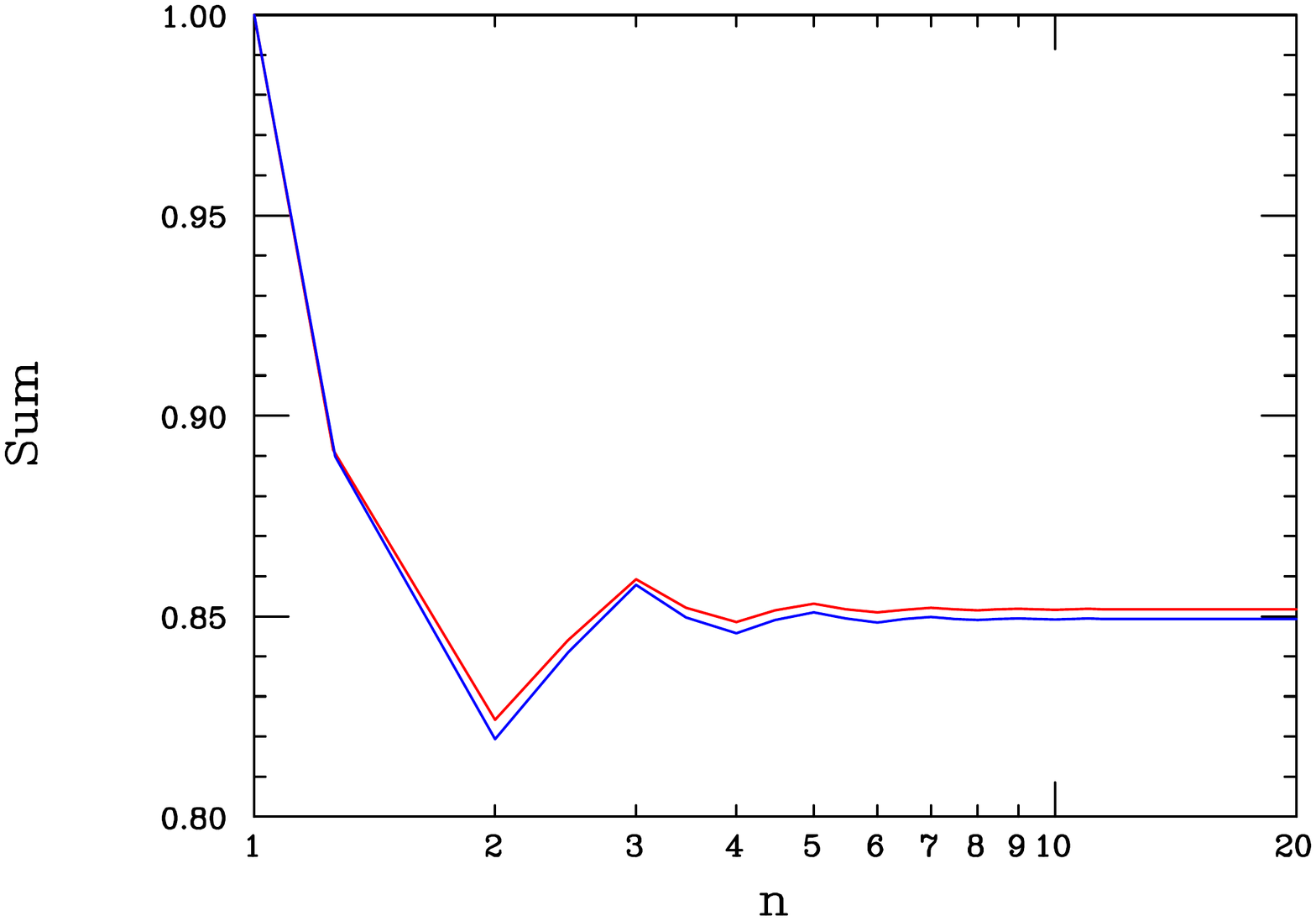}
\hspace {-2.5cm}
\includegraphics[width=3.5in,angle=90]{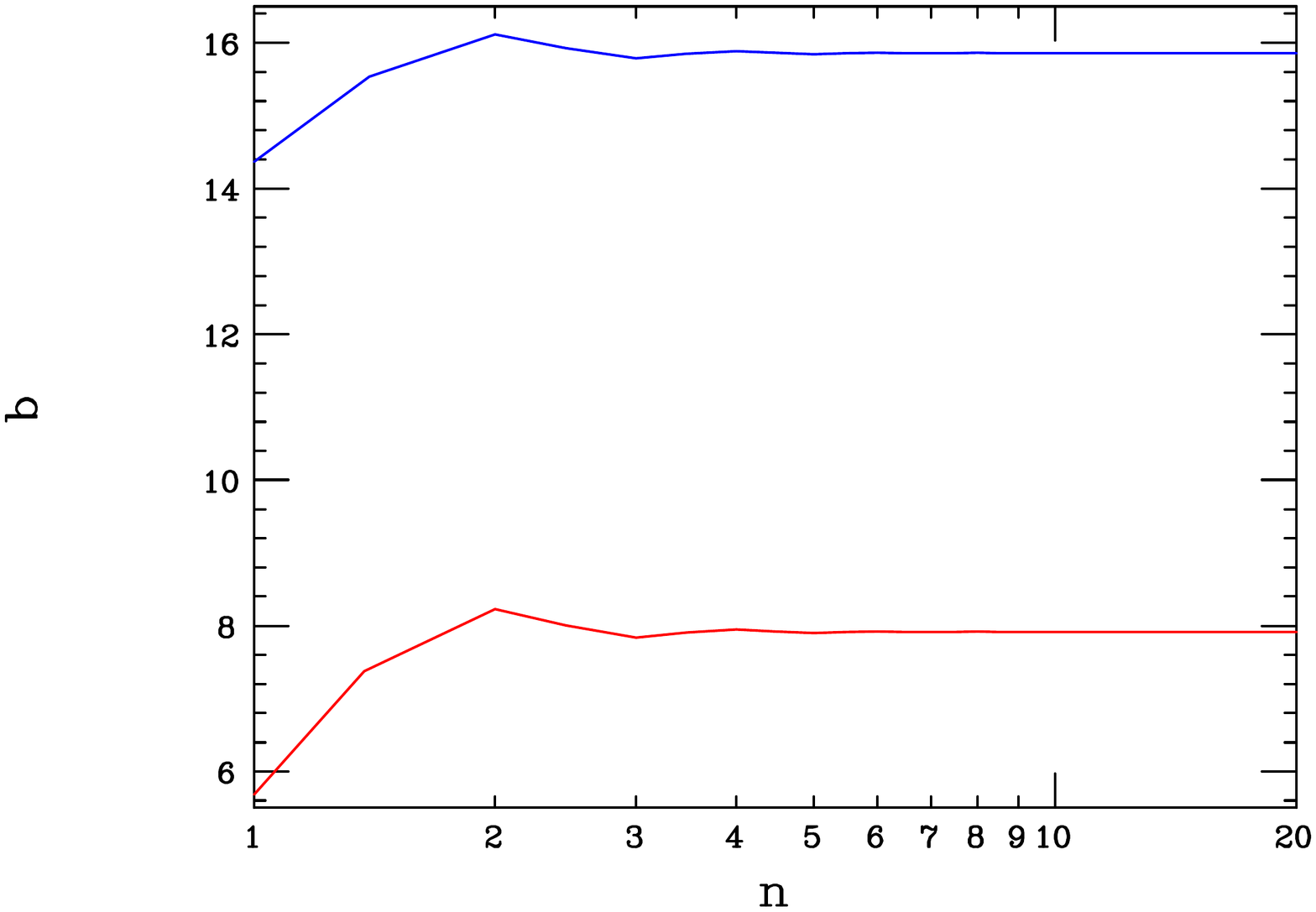}}
\vspace*{-1.50cm}
\caption{Left: Values of the quantity `Sum' appearing in the DM-electron SI scattering cross section as a function of the gauge KK tower number, $n$, 
included in the sum,  as described in the text; the upper(lower) curve corresponds to the case of BM1(BM2).  Right:  The value of the quantity $b$, as defined in the text, for BM1 
(upper curve) and for BM2 (lower curve).}
\label{dm1}
\end{figure}

Given a fixed set of model parameters, the couplings of the KK scalar tower states to the gauge KK states, $S_n^\dagger S_mV_i$,  can be calculated from their associated wavefunctions 
by performing the integrals
\begin{equation} 
g_Dc_{mn}^i  \sim g_{5D} \int_0^{\pi R} ~s_n(y)s_m(y)v_i(y)~dy\,
\end{equation}
where we can define the dimensionless 4-D coupling, $g_D$, in terms of the lowest mode states in each tower. Note that once $\delta_{A,S}$ and the product $g_D\epsilon_1$ are 
specified, all physically relevant observables become calculable in term of the value of $R$ or in terms of an overall mass scale, \eg, $m_{S,V_1}$. For example, the SI cross section for 
DM scattering off electrons is given by 
\begin{equation} 
\sigma_e = \frac{4\alpha \mu^2 g_D^2 \epsilon_1^2}{(m^V_1)^4} ~\Bigg[\sum_n ~(-1)^{n+1} \frac{\epsilon_n}{\epsilon_1}~c_{11}^n ~\frac{m_{V_1}^2}{m_{V_n}^2}\Bigg]^2\,
\end{equation}
where $\mu=m_em_{DM}/(m_e+m_{DM})$ is the reduced mass $\sim m_e$  for the DM mass values of interest to us. Numerically this yields the result 
\begin{equation} 
\sigma_e \simeq 3.0\cdot 10^{-40} \rm{cm}^2 ~\Big(\frac{100~ \rm{MeV}}{m_{V_1}}\Big)^4 ~\Big(\frac{g_D\epsilon_1}{10^{-4}}\Big)^2 ~\times \rm{Sum}\,
\end{equation}
whereby the quantity `Sum' represents the squared KK summation of the previous expression which is expected to be $\sim O(1)$  as the series converges very rapidly. Note that here `Sum' 
isolates the difference between the prediction of the 5-D scenario and the 4-D case. For representative parameter values, \eg,  SuperCDMS is likely to be able to probe this range of cross sections 
in the future but now they lie a few orders of magnitude below the current constraints. The calculation of the thermal DM annihilation cross section into, \eg,  final state electrons can be expressed 
in a similar fashion by writing $\sigma v_{rel}=\tilde bv_{rel}^2$, where the detailed kinematic information, including the sub-leading terms in the velocities, and (away from any resonances 
for simplicity) is contained in the parameter $\tilde b$ which in the limit of a zero electron mass is given by
\begin{equation} 
\tilde b=\frac{g_D^2 e^2 \epsilon_1^2}{192\pi m_{DM}^2} ~\frac{\gamma^4}{\gamma^2-1}~ \sum_{n,m} (-1)^{n+m} \Bigg[\frac{(\epsilon_n\epsilon_m/\epsilon_1^2)~c_{11}^n c_{11}^m}
{(\gamma^2-r_n)(\gamma^2-r_m)}\Bigg] \,
\end{equation}
where here the double sum is over the gauge KK tower states, $\gamma^2=s/4m_{DM}^2$ is the usual kinematic factor determined by the DM velocities employing the standard 
Mandelstam variable and  $r_n=m_{V_n}^2/4m_{DM}^2$. We will assume the freeze-out temperature to be $x_F=m_{DM}/T\simeq 20$ so that at freeze-out 
$<v_{rel}^2>\simeq 0.3$ in numerical estimates. For the benchmark models that will be discussed below we not only have  $2m_{S_1} > m_{V_1}$ but also that 
$2m_{S_1}$ is significantly below $m^V_2$ implying that the thermal DM annihilation cross section is dominated by phase space regions far from any of the narrow $s$-channel 
KK resonances. To go further we need to choose some specific benchmark models (BM) forcing us into some  
particular parameter choices. Here we give two examples both of which have $\delta_A=0.5$. For BM1, we take $m_{V_1}/m_{DM}=0.8$ implying $\delta_S\simeq 2.38$, while 
for BM2, we assume that $m_{V_1}/m_{DM}=0.6$ implying $\delta_S\simeq 6.03$. Apart from the overall mass scale set by $R^{-1}$, these quantities determine the 
complete model phenomenology.  In the left panel of Fig.~\ref{dm1} we find the value of the quantity `Sum' defined above for our two benchmark points as a function of the 
number of contributing gauge KK tower states $n$. Here we see that ($i$) the results for these two BM points are essentially identical, ($ii$) the KK summation converges very 
rapidly, roughly by the time the $n\sim 5$ KK state is reached. ($iii$) The value of this sum is less than unity due to the destructive interference among the gauge KK 
exchanges, \ie, `Sum' $\simeq 0.852(0.849)$ for BM1(BM2). This means that the entire KK tower above the lowest level makes only a $\sim 7\%$ contribution to the amplitude. 
Finally, ($iv$) we see that the 4-D and 5-D predictions are numerically quite close. It is important to emphasize the very rapid convergence of these sums and the essentially negligible 
contributions of the higher KK states here. The right panel of Fig.~\ref{dm1} shows the values of a quantity $b$ for both BM points; in this Figure we rescaled the quantity $\tilde b$ 
above by an overall factor so that this quantity $b$  as shown here is dimensionless and is roughly $O(10)$:  
\begin{equation}
\tilde b =b ~\Bigg[\frac{g_D^2e^2\epsilon_1^2}{m_{DM}^2 \rm {(GeV^2})}\Bigg] ~10^{-20} \rm{cm^3 s^{-1}}\,.
\end{equation}
In this panel we further see that BM1(BM2)  leads to a value of $b \simeq 7.9(15.9)$ which differ by roughly a factor of $\sim 2$ due to the BM mass spectrum and various coupling 
variations. It is easy to see that for $g_D\epsilon_1 \sim 10^{-4}$ and $m_{DM} \sim 10-100$ MeV we can straightforwardly obtain a thermal cross section of $\sim 9\cdot 10^{-26}~\rm{cm^3s^{-1}}$ 
as needed to reproduce the observed relic density for light complex DM masses.  We again emphasize the very rapid convergence of these KK sums and the essentially 
negligible contributions of the higher KK states beyond $n\sim 5$ for both these observables.

It is clear that using these two observables alone it will be quite difficult to differentiate the 5-D from the 4-D models; in fact, when $\lambda<0.5$ we only have ME signatures to accomplish this. 
For a true separation of these two possibilities clearly we must produce some of the KK modes on-shell. In the $\lambda <1/2$  case the cleanest approach is to employ the $\gamma$+ME 
final state in either meson decays or in $e^+e^-$ annihilation where multiple photon recoil peaks may be observable 
associated with the production of the different DP KK states. Other techniques employing, \eg, DP tower direct production in fixed target collisions may be also be helpful but in this case 
the signals that are useful in separating the 4-D from 5-D scenarios are much more subtle and will depend upon detailed knowledge of the anticipated rates and associated distributions with 
high precision. When $\lambda>0.5$,  as in the case of our BM points, the 5-D and 4-D signatures are much easier to differentiate since the heavier KK production leads to visible cascade 
decays which can be rather complex. Of course, by construction, for both BM points, $S_1$ and $S_1^\dagger$ are stable states 
forming the DM while $V_1$ decays only into SM final states as the decay $V_1\to S_1^\dagger S_1$ is kinematically forbidden. Furthermore, $V_2$ essentially only decays into 
$S_1^\dagger S_1$ since $g_D^2>>(e\epsilon_1)^2$.  In this 5-D model, the $V_1$ acts similar to the 4-D DP decaying to only SM states while the $V_2$ acts like the 4-D model 
where the DP decays only to DM. In a similar fashion, the decay $S_2\to S_1V_1$ occurs with a $\sim$100\% branching fraction.  The decays of 
the higher KK states are found to be somewhat sensitive to the BM choice due to their differences in couplings and phase space although the gauge KK masses are the same for both BMs. 
\begin{table}
\centering
\caption{Branching fractions for the various decay modes in per cent for the next highest gauge and scalar KK states in both BM scenarios as discussed in the text.}
\vspace{0.3cm}
\begin{tabular}{|l|c|c|c|} \hline
~~~Process             &    BF(BM1)    &   BF(BM2)   \\
\hline
$S_3\to V_2S_1$  &  1.20     &  0.62 \\
$S_3\to V_1S_1$  &   5.10    &  1.78   \\
$S_3\to V_1S_2$  &   93.7     &  97.6   \\
$V_3\to S_1^\dagger S_1$  &  74.9     &   97.3    \\
$V_3\to S_1^\dagger S_2$+h.c.  &   25.1     &  2.71   \\
$V_4\to S_1^\dagger S_1$ &  45.9     & 39.5      \\
$V_4\to S_1^\dagger S_2$+h.c. &   51.5    & 18.9       \\
$V_4\to S_2^\dagger S_2$  &  1.67      & 38.8    \\
$V_4\to S_3^\dagger S_1$+h.c.  &  0.95     &  2.81     \\
\hline
\end{tabular}
\label{decay1}
\end{table}
In Table~\ref{decay1} we see that there can be quite significant differences in how the various KK states decay based on the small differences in masses and the variations in the $c_{nm}^i$ 
couplings. Searches for these more massive KK states will be somewhat influenced by these parametric variations. The fact that these two BMs can show such differences suggests 
that even greater variations are likely possible as we scan over the full parameter space.  As noted, once decays of these light KKs into other dark sector states are kinematically 
allowed the corresponding lifetimes are generally controlled by the coupling factors $\sim g_D^2 \times ~O(1)$  so that such decays are quite rapid. Of course the lightest KK gauge 
state, which decays to SM fields via $(e\epsilon_1)^2$ can also be long-lived as has been often discussed in the literature for the 4-D case with typical $c\tau$ 
values of order 100 $\mu$m for $\epsilon_1 \sim 10^{-4}$ and masses of $\sim 100$ MeV. As we progress up the various KK towers, decay widths will increase due to the usual opening 
of phase space and overall mass factors although in most cases these will be somewhat compensated for by the shrinking values of the relevant parameters $c_{nm}^i$ and the compression 
of phase space for some decay modes due to near mass degeneracies.

\section{Conclusions}

ED extensions of known scenarios can lead to additional model building flexibility, address some of the issues that arise in 4-D and can lead to new interesting phenomenology. This is 
particularly useful in the case of DM where our limited knowledge requires all accessible avenues be explored. Hopefully one of these avenues will lead us to the discovery of DM.

\section*{Acknowledgments}

This work was supported by the Department of Energy, Contract DE-AC02-76SF00515 as SLAC-PUB-17245. I would like to thank J.L Hewett for lengthy and illuminating discussions on 
this topic.

\end{document}